\documentclass[twocolumn,showpacs,preprintnumbers,amsmath,amssymb]{revtex4}

\usepackage{graphicx}
\usepackage{dcolumn}
\usepackage{bm}

\newcommand{\avk}{<\!\!k\!\!>}
\newcommand{\avkk}{<\!\!k^2\!\!>}
\newcommand{\avks}{<\!\!k\!\!>_{k^*}}
\newcommand{\avkks}{<\!\!k^2\!\!>_{k^*}}

\begin{document}

\preprint{APS/123-QED}

\title{2-Peak and 3-Peak Optimal Complex Networks}

\author{Andr\'{e} X. C. N. Valente$^{1}$}
\email{andre@deas.harvard.edu}

\author{Abhijit Sarkar$^{2}$}
\author{ Howard A. Stone$^{1}$}
 \affiliation{
$^{1}$Division of Engineering and Applied Sciences\\
Harvard University, Cambridge, MA 02138, U.S.A.\\
$^{2}$Department of Physics and Department of Chemistry and Chemical Biology\\
Harvard University, Cambridge, MA 02138, U.S.A.\\
}%


\date{\today}

\begin{abstract}
A central issue in complex networks is tolerance to random failures and intentional attacks. Current literature emphasizes the dichotomy between networks with a power-law node connectivity distribution, which are robust to random failures but fragile to targeted attacks, versus networks with an exponentially decaying connectivity distribution, which are less tolerant to failures but more resilient to attacks.
We prove analytically that the optimal network configuration under a classic measure of robustness is altogether different from both of the above: in all cases, failure and/or attack, there are no more than three distinct node connectivities in the optimal network.
\end{abstract}

\maketitle

An outstanding issue in systems as diverse as power supply \cite{SmallWorld}, transportation \cite{AmaralAirports}, communication \cite{Faloutsos,BarabAttack}, gene \cite{Metabolic1,GeneCircuits}, metabolic \cite{Metabolic1} and ecological \cite{Ecological} networks is tolerance to component breakdown.
Recent research on these and other complex networks has focused on abstracting from the details intrinsic to each of the systems and represents them in a unified way as a network of nodes connected by links \cite{StrogatzRev,BarabRev,Wang,Motifs,LifesPyramid}.

Breakdown in a complex network is represented by the removal of nodes \cite{LinksCase} and robustness refers to the ability of the surviving nodes to remain, as much as possible, interconnected.
Two cases of fundamental interest are the the random removal of nodes, which simulates random failures of individual elements, and the removal of the most connected nodes, which simulates a targeted attack aimed at crippling a network \cite{BarabAttack,BarabRev,Epstein}.

We report the results of an analysis of the problem of designing networks to be robust against random failures and deliberate attacks and obtain a constructive proof of the most robust network architecture.
The network configurations we find are remarkably simple to describe qualitatively: They are characterized by the presence of at most three distinct node connectivities in the network.

A simple but essential measure that is used to capture the structure of a network is the node degree distribution:
 The degree, or connectivity, of a node is the number of links emanating from it, while the degree distribution gives the probability that a randomly chosen node has a given degree.
Our analysis applies to the class of networks, known as generalized random graphs \cite{GenFunc, Molloy1}, which are random in every respect other than in their specified degree distribution.
In generalized random graphs, potential node degree correlations in the network~\cite{Correl} are ignored. In particular, in the limit of a large number of nodes, the fraction of nodes forming loops of a given size goes to zero \cite{GenFunc,CohenRand}.

The robustness criterion we apply is the presence (or absence) of a giant connected component in the network, defined as a connected cluster of nodes whose size scales linearly with the network size $N$ (the total number of nodes) \cite{Molloy1}. 
This analysis pertains to the case where N tends to infinity. In particular, note that, given a degree distribution, N must be large enough so that any degree with nonzero probability in the distribution is present in a statistically significant number of nodes in the network.  
In a phase (or percolation \cite{BarabRev}) transition, the giant connected component disappears under the removal of more than some critical fraction of the network nodes \cite{BarabRev,GenFunc,Molloy1,NewmanPerc}, with the size of the largest connected cluster then scaling only proportionally to $\log N$ \cite{Molloy1}.
Naturally, the percolation threshold depends on whether the nodes are removed randomly (random failure mode) or whether the most connected ones are chosen for removal first (attack mode): we denote the two distinct thresholds by $f_r$ and $f_a$, respectively. 
This is a topological robustness criterion, as it does not take into account potential dynamical effects subsequent to the removal of a node. For an example of interplay between dynamics and robustness see \cite{Watts,Motter,Moreno} and for other possible topological measures of robustness see \cite{BarabRev,Epstein}.

With the above robustness criterion in mind, we define the optimization problem as follows: Given a fixed average number of links per node, we determine the node degree distribution that maximizes the percolation threshold. 
An equivalent view of the problem is that of finding the degree distribution that minimizes the average number of links per node while still satisfying a minimal percolation threshold, specified apriori, as a desired robustness condition.
We shall use these two perspectives interchangeably in the rest of the letter.
As a final constraint on the optimization problem, we require that the degree $k$ of every node in the network satisfies $0\!<k_\ell\leq k\!\leq\! k_m$, where the minimum $k_\ell$ and maximum $k_m$ allowed degrees are considered given and fixed. 
We analyze three variations on the optimization problem: Optimization against only random failures, optimization against an intentional attack and, finally, the most practically relevant case, optimization of a network against both random failures and intentional attacks.

We start with the case of optimization against random failures.
Let $\avk$ denote the average degree in the network and $\avkk$ denote the average squared degree.
Then, the percolation threshold $f_r$ for a network with a given degree distribution is given by \cite{CohenRand,NewmanPerc}
\begin{equation} \label{PercCritRand}
f_r=1-\frac{1}{\frac{<k^2>}{<k>}-1}\quad.
\end{equation}

Clearly, the optimal degree distribution, which maximizes $f_r$, maximizes $\avkk$ for a fixed $\avk$. Now, the distribution that maximizes $\avkk$ for a fixed $\avk$ is the one where all the nodes have either degree $k_\ell$ or degree $k_m$, the two extreme degrees.
We conclude that the optimal degree distribution can be nonzero only at $k_\ell$ and $k_m$, with the choice for the exact partition of the distribution between these two degrees being determined by the value of $\avk$.
We call such a distribution with only two nonzero values a ``2-peak distribution''.
Under this optimal 2-peak distribution, $f_r$ can be expressed as
\begin{equation}
f_r=1-\frac{\avk}{-k_m k_\ell+\avk(k_m+k_\ell-1)}\quad.
\end{equation}

Fig. 1 shows how two archetypical real networks fair by comparison: The electrical power grid of the western United States, which is a network with an exponentially decaying degree distribution and the internet router network, which is a network with a power-law degree distribution.

\begin{figure}
\scalebox{.32}{\includegraphics{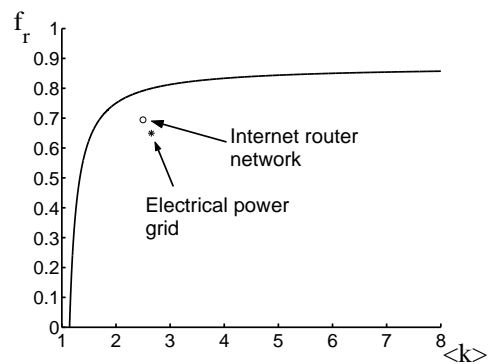}} \label{f-k-rand}
\caption{Optimization against random failures. Plot of $f_r$ the critical fraction of randomly deleted nodes where the giant connected component vanishes versus $\avk$ the average number of links per node in the network, under an optimally chosen degree distribution - in the sense that it maximizes $f_r$ for a given $\avk$. A permissible minimum $k_\ell\!=\!1$ and maximum $k_m\!=\!8$ node degree constraint was also imposed. For $\avk$ below $8/7$ there is no percolation in the network even with no node failures. Maximum robustness occurs when all the nodes have $k_m$ links, in which case the percolation transition occurs at $f_r\!=\!6/7$. For comparison purposes $f_r$ and $\avk$ for two real networks are plotted: * - Western United States electrical power grid, an exponential network. o - Internet router network, a power-law network. For the power grid $(k_\ell,k_m)\!=\!(1,19)$ and for the internet $(k_\ell,k_m)\!=\!(1,20)$ \cite{cutoff}. The values of $f_r$, $\avk$, $k_\ell$ and $k_m$ for these real networks were computed from data in references \cite{Faloutsos,ScaleFree}.
}
\end{figure}

We now analyze optimization against intentional attacks.
In this case nodes are removed sequentially in descending order according to their degree, starting with the most connected node.
Let $f_a$ be the minimal desired percolation threshold against attack.
Using the generating function formalism \cite{GenFunc,NewmanPerc}, we express in a useful form the condition for percolation when the fraction $f_a$ of the most connected nodes is deleted \cite{NewmanPerc,CohenAttack}.
Let us visualize an attack on a network as described in Fig. 2.
\begin{figure}
\label{White-Black-Fig}
\scalebox{.10}{\includegraphics{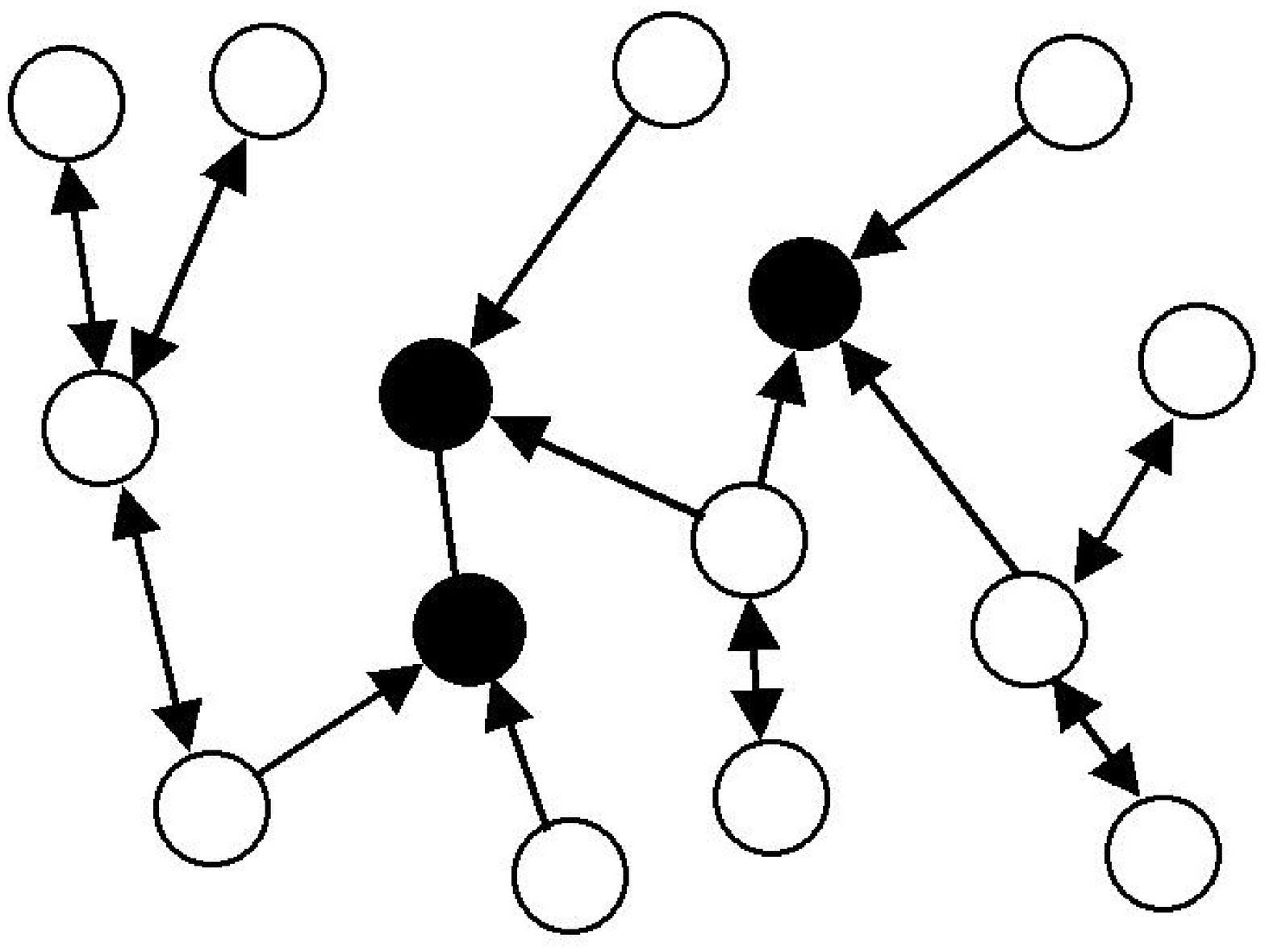}}
\caption{Network under intentional attack. Functional nodes are white and nodes destroyed under attack are black. A link between white nodes can be traversed both ways. The non-functionality of a black node is characterized by having links to it become one-way incoming links only, i.e., such links work in the direction leading into the black node, but not going out of the black node (note: a link between two black nodes cannot be transversed either way).}
\end{figure}
For the purpose of determining the existence of a giant component, this view of the change in connectivity is entirely equivalent to that of deleting the attacked nodes and all links emanating from them. However, the generating functions for this network are now elementary. Let $G_0(x)$ be the generating function associated with the distribution of outgoing links of a node picked at random and let $G_1(x)$ be the generating function associated with the distribution of remaining outgoing links of a node arrived at by following a link emanating from a white node. Then, with $p_k$ denoting the probability that a node picked at random has degree $k$ and with $q_k$ standing for the fraction of nodes of degree $k$ that are white, we have
\begin{align}  
  &G_0(x)=\sum _{k=k_\ell}^{k_m} p_k\, (1\!-\!q_k) +\sum_{k=k_\ell}^{k_m} p_k\, q_k \, x^k \quad,\\
  &G_1(x)=\frac{\sum _{k=k_\ell}^{k_m} k\, p_k\, (1\!-\!q_k)+\sum_{k=k_\ell}^{k_m} k\, p_k \, q_k \, x^{k-1}}{\sum_{k=k_\ell}^{k_m} k \, p_k} \quad.
\end{align}
Thus, using standard methods  \cite{NewmanPerc, GenFunc}, the percolation condition is
\begin{align}
& & G_1^{\,\prime}(1)&\geq1 \qquad\Rightarrow & \frac{\sum_{k=k_\ell}^{k_m} k^2\, p_k \, q_k -\sum_{k=k_\ell}^{k_m} k\, p_k \, q_k }{\sum_{k=k_\ell}^{k_m} k \, p_k} &\geq 1 \quad. \label{AttCond}
\end{align}

Under attack, the function $q_k$ takes the form 
\begin{equation}
q_k=\begin{cases}
0 &\text{for $k\!>\!k^*$}\\
q_{k^*} &\text{for $k\!=\!k^*$}\\
1 &\text{for $k\!<\!k^*$}
\end{cases}\qquad,
\end{equation}
where $k^*$ is the largest degree a white node may have after the fraction $f_a$ of the most connected nodes in the network has been darkened.
Given this form of $q_k$, we express the percolation condition (\ref{AttCond}) as
\begin{equation}
\frac{\avkks-\avks}{\avk}\geq 1 \quad. \label{AttCond2}
\end{equation}
This shorthand notation serves to highlight the fact that the numerator consists of the averages of $k^2$ and $k$ associated with the original $p_k$ distribution modified by transforming all nodes with degree above some $k^*$ (including some fraction $(1-q_{k^*})$ of the nodes of degree $k^*$) into nodes of degree zero.

We now argue that the optimal distribution against attack is also a 2-peak distribution, namely, one where $p_k$ can be nonzero only at $k_\ell$ and at $k^*$. 
Let us consider a graph with an arbitrary $p_k$ degree distribution and with a fraction $f_a$ of its nodes destroyed under attack. Moving all the probability in the $k\!>\!k^*$ region to the point $k^*$ affects neither $\avkks$ nor $\avks$ and yet it decreases $\avk$. In other words, it widens the inequality (\ref{AttCond2}) while decreasing the average number of links used. We conclude that, in the optimal degree distribution, $p_k\!=\!0$ for $k\!>\!k^*$. Now let us consider the probability in the region $k\!<\!k^*$. By an argument analogous to the one used for the random failures case, $\avkks$ can be maximized while keeping $\avks$ and $\avk$ fixed by concentrating all the probability in the region $k\!<\!k^*$ at the extreme values $k_\ell$ and $k^*$. Therefore we conclude that the optimal distribution against intentional attacks is also a 2-peak distribution with the entire probability concentrated at $k_\ell$ and at $k^*$.
The optimization problem is therefore reduced to finding the $k^*\!\in\!\{ k_\ell,...,k_m\}$ and $p_{k^*}\!\in\![0,1]$ values that minimize $\avk$ subject to condition~(\ref{AttCond2}).
This is further simplified by noting that, at each candidate $k^*$, only the $p_{k^*}$ value that yields equality in condition (\ref{AttCond2}) needs to be checked (excluding the trivial case in which the minimal distribution where all the nodes have degree $k_\ell$ suffices to satisfy the percolation requirement). 
The different $k^*$ candidates are therefore trivially checked numerically.
As an aside, note that, unlike in the random failures case, in the attack case the presence of a $k_m$ maximum allowed degree is not strictly necessary, since the analysis will always yield a finite $k^*$. 

Fig. 3a shows $f_a$ as a function of $\avk$ for an optimal network. The values of $f_a$ and $\avk$ for the electrical power grid and internet router networks are provided for comparison. The characterization of the optimal network robust to a given $f_a$ is completed by giving the values of $k^*$ and $p_{k^{*}}$, shown in Fig. 3b and 3c, respectively.

\begin{figure}
\scalebox{.42}{\includegraphics{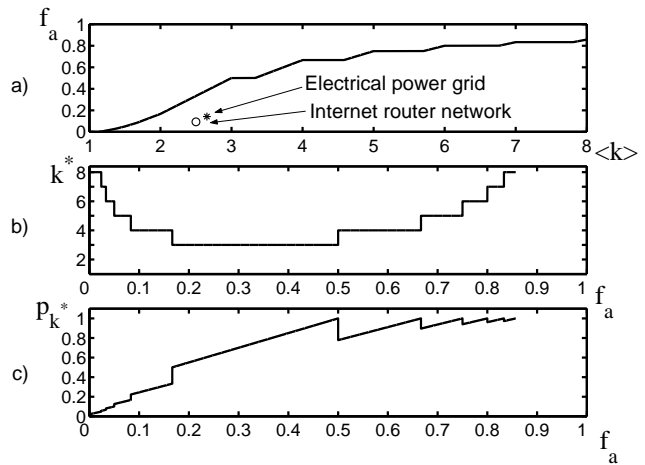}} \label{AttackPic}
\caption{
Optimization against intentional attacks. 
(a) Analogous plot to Fig. 1, except now the nodes are deleted under attack mode, meaning the nodes are removed sequentially in descending order according to their degree, starting with the most connected node. Again, we take $(k_\ell,k_m)\!=\!(1,8)$.  * - Western United States electrical power grid. o - Internet router network.
(b) $k^{*}$, one of the two node degrees present (the other being $k_\ell$) in the associated 2-peak optimal degree distribution against attack.
(c) $p_{k^*}$, the fraction of nodes with degree $k^{*}$ in the optimal degree distribution. The discontinuity in the variables is a consequence of the discreteness of $k^*$.}
\end{figure}

We now find the degree distribution for a network that, while minimizing the average number of links per node used, percolates under the loss of a fraction $f_a$ of its nodes under attack and percolates under the loss of a fraction $f_r$ of its nodes under random failure; we demand that the network satisfy the two percolation conditions separately, that is, the specification is not that the network percolates under a combined fractional loss of $f_a\!+\!f_r$.
In this case we argue that the optimal degree distribution can be a 3-peak distribution - it's nodes have only one of three possible degrees: $k_\ell$, $k_m$ and an in  between degree $k^*$, where the previous definition of $k^*$ continues to hold.
Let us consider an arbitrary degree distribution $p_k$. Moving the entire probability in the region $k\!<\!k^*$ to the extremes $k_\ell$ and $k^*$ while keeping $\avk$ and $\avks$ constant (same condition), has the effect of maximizing $\avkk$ and $\avkks$. This benefits both the robustness against attacks, equation~(\ref{AttCond2}), and the robustness against random failures, equation~(\ref{PercCritRand}). We conclude that there are no nodes with degree in between $k_\ell$ and $k^*$ in the optimal degree distribution. We now turn to the probability in the region $k\!>\!k^*$. As far as robustness against random failures, equation~(\ref{PercCritRand}), moving all the probability to the extremes $k^*$ and $k_m$ while keeping $\avk$ constant, maximizes robustness by maximizing $\avkk$. On the other hand, robustness against attack, equation~(\ref{AttCond2}), is not affected by the placement of the probability in the region $k>k^*$ as long as $\avk$ is kept constant, since $\avkks$ and $\avks$ are not affected. We conclude that there are no nodes with degree in between $k^*$ and $k_m$ in the optimal degree distribution. Therefore, we have shown that the optimal distribution can have at most three distinct degrees.
The exact qualitative form of the distribution depends on the combination of $f_r$ and $f_a$. The different possibilities are outlined in Fig. 4a, using a color coding scheme. Fig. 4b shows the $\avk$ of the optimal network associated with each given $(f_r,f_a)$ pair. 
We note that numerically only a limited number of potential solutions need to be checked to find the optimal architecture. This limited checking follows because, for the cases in which the solution is determined by a combination of $f_a$ and $f_r$, it can be shown that if the optimal solution is a 2-peak distribution, then equality holds in (\ref{AttCond2}), while if the optimal solution is a 3-peak distribution, equality (\ref{PercCritRand}) also holds in addition to (\ref{AttCond2}). Therefore the potentially optimal values of $p_{k^*}$ and $p_{k_m}$ can always be expressed in terms of the candidate $k^*$.        

\begin{figure}
\scalebox{.48}{\includegraphics{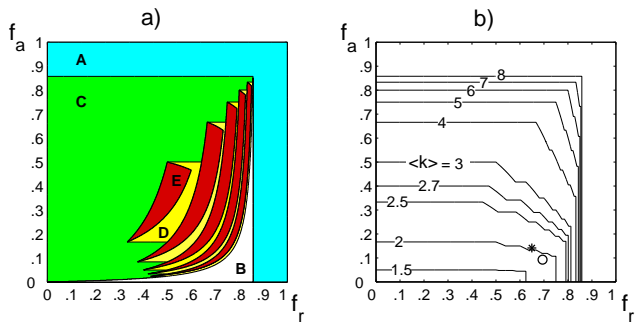}} \label{SimulPic}
\caption{Simultaneous optimization against intentional attacks and random failures. We take $(k_\ell,k_m)\!=\!(1,8)$.
To each combination of desired minimal network percolation thresholds, $f_a$ under attack and $f_r$ under random failures, corresponds an optimal network, i.e., one that also minimizes $\avk$. 
(a) These optimal networks can be divided into different qualitative classes, illustrated using different colors: 
A - Robustness to these $(f_a, f_r)$ pairs is not attainable due to the $k_m$ constraint.
B - $f_r$ is the limiting constraint. There are two node degrees present in these networks, $k_\ell$ and $k_m$.
C - $f_a$ is the limiting constraint. There are at most two distinct node degrees in these networks, $k_\ell$ and $k^*$.
D - Both $f_a$ and $f_r$ affect the optimal degree distribution. These networks still have just two distinct node degrees,  $k_\ell$ and $k^*$ (i.e., the potential third degree, $k_m$, turns out to have zero frequency).
E - As in D, both $f_a$ and $f_r$ affect the optimal degree distribution but there are now three distinct node degrees in the network, $k_{\ell}$, $k^*$ and $k_m$.
(b) Contour plot of $\avk$ for the optimal networks. The $\avk\,=k_m\!=\!8$ contour represents the maximum achievable robustness.
For comparison, the $(f_r,f_a)$ robustness thresholds of two real networks were plotted: * - Western United States power grid (exponential network), o - Internet router (power-law network).
For the power grid $\avk\,=\!2.7$ and for the internet $\avk\,=\!2.5$. Note how the points fall below the respective optimal $\avk$ contours.}
\end{figure}

In this letter we have shown that the network configurations that maximize the percolation threshold under attack and/or random failures have at most three distinct node degrees.
From a practical point of view, both engineered and naturally occurring networks have a diversity of factors influencing and constraining their ultimate configuration.
Nonetheless, the optimal configurations we present provide a yardstick against which the robustness of real networks can be compared and act as an intuitive guide for network-robustness engineering.\\

A.S. deeply appreciates the guidance and support of Professor David Nelson and Professor X. Sunney Xie, as well as the financial support of NSF (DMR-0231631) and HMRL (DMR-0213805).

\bibliographystyle{plain}

\begin{thebibliography}{99}
\bibitem{LinksCase}
Some problems are best modeled with the removal of links rather than nodes. The mathematical analysis for such cases parallels closely that for the case of node removal. See for example \cite{NewmanEpidemics}.
\bibitem{cutoff}
Although $k_m$ for the power grid was 19, only 1.5\% of the nodes had a degree above 8. Similarly, only 3\% of the internet nodes had a degree above 8. In face of this, we considered that imposing a maximum of $k_m\!=\!8$ in our optimal peaked networks would make the comparison fairer. Had we imposed $k_m\!=\!19$ or 20, the robustness difference between our peaked networks and the internet and power grid networks would only increase.
\end{thebibliography}

\end{document}